\documentstyle[12pt,epic,eepic]{article}

\makeatletter
	\def\@cite#1#2{\leavevmode\hbox{$^{\mbox{\the\scriptfont0 #1}}$}}
\makeatother

     \def\sR{{\scriptscriptstyle R}}
     \def\sD{{\scriptscriptstyle D}}

     \def\11{1\mbox{\hspace{-0.9ex}}1}
     \def\tr{\mbox{tr}}

\makeatletter
\long\def\@makecaption#1#2{%
   \vskip 10\p@
   \setbox\@tempboxa\hbox{#1\ \ #2}%
   \ifdim \wd\@tempboxa >\hsize
	#1\ \ #2\par		%
      \else
	\hbox to\hsize{\hfil\box\@tempboxa\hfil}%
   \fi}
\def\fnum@figure{Fig. \thefigure}
\makeatother

\begin{document}

\begin{flushright}
\begin{tabular}{l}
HUPD-9522 \hspace{0.5cm}\\
December 1995
\end{tabular}
\end{flushright}

\vspace{1cm}

\begin{center}
\bf
Curvature Induced Phase Transition in a Four-Fermion Theory\\[2mm]
Using the Weak Curvature Expansion\\[1.5cm]
\normalsize
Tomohiro~Inagaki
\footnote{e-mail: inagaki@theo.phys.sci.hiroshima-u.ac.jp}
\\[8mm]
{\it
Department of Physics, Hiroshima University, \\
Higashi-Hiroshima, Hiroshima 739, Japan \\[3cm]
}
\end{center}

\begin{abstract}
\noindent
Curvature induced phase transition is thoroughly investigated in a
four-fermion theory with $N$ components of fermions for arbitrary space-time
dimensions $(2 \leq D < 4)$.
We adopt the $1/N$ expansion method and calculate the effective potential for
a composite operator $\bar{\psi}\psi$.
The resulting effective potential is expanded asymptotically
in terms of the space-time curvature $R$ by using the Riemann normal
coordinate.
We assume that the space-time curves slowly and keep only terms
independent of $R$ and terms linear in $R$.
Evaluating the effective potential it is found that the first-order phase
transition is caused and the broken chiral symmetry is restored for a large
positive curvature.
In the space-time with a negative curvature the chiral symmetry is
broken down even if the coupling constant of the four-fermion interaction
is sufficiently small.
We present the behavior of the dynamically generated fermion mass.
The critical curvature, $R_{cr}$, which divides the symmetric and asymmetric
phases is obtained analytically as a function of the space-time dimension $D$.
At the four-dimensional limit our result $R_{cr}$ agrees with the exact
results known in de Sitter space and Einstein universe.
\end{abstract}

\newpage

\renewcommand{\thesubsection}{\arabic{subsection}}
\renewcommand{\thesubsubsection}
	{\arabic{subsection}.\arabic{subsubsection}}
\baselineskip=22pt

\subsection{Introduction}

\hspace*{\parindent}
After the pioneering work by Y.~Nambu and G.~Jona-Lasinio,\cite{NJL}
the idea of dynamical symmetry breaking has played a decisive role
in modern particle physics.
The idea was introduced to discuss the chiral symmetry breaking in QCD.
S.~Weinberg and L.Susskind apply the idea to the electroweak
symmetry breaking through the technicolor scenario.\cite{TC}
There is a possibility that the gauge symmetry of the grand unified theory
is broken down dynamically.

It is expected that the primary symmetry of unified theories is
broken down at the early universe to yield lower level
theories which describe phenomena at a lower energy scale.
The models of unified theories may be tested in critical phenomena
at the early universe.
As a first step to investigate the unified theories much interest has
been taken in clarifying the mechanism of the spontaneous symmetry
breaking under the circumstance of the early universe.
One of the possible mechanism is the dynamical symmetry breaking which is
caused by the non-vanishing vacuum expectation value of the composite
operator $\bar{psi}\psi$ which is constructed by a fermion and an anti-fermion.

In the early universe it is not adequate to neglect the effects of the
curvature, temperature and density.\cite{IKM}
In the present paper a four-fermion theory is studied as a prototype model
of the dynamical symmetry breaking and we study the dynamical symmetry
breaking caused by the curvature effect in general curved space-time.
We assume that the space-time curvature is small and evaluate the
effective potential of the theory in arbitrary dimensions, $2 \leq D < 4$.
In the GUT era it seems plausible to treat the space-time geometry
classically.

Many works have been done in this field.
In four dimensions Nambu-Jona-Lasinio model is studied by using the weak
curvature expansion.
In the model the chiral symmetry is restored for a large positive curvature
and the phase transition is of the first order.\cite{CURV}
In the $D$-dimensional de Sitter space and Einstein universe
it is found to exhibit the symmetry restoration through the second order
phase transition.${}^{5 \sim 10}$
I.~Sachs discuss the validity of the weak curvature expansion in four
dimensions.\cite{SACH}

We make a systematic study of the four-fermion theory
in arbitrary dimensions using the weak curvature expansion
and show the validity of the weak
curvature expansion for a large positive curvature.
In Sec.2 we introduce a four-fermion theory in curved space-time.
We employ the $1/N$ expansion method to obtain the effective potential.
In Sec.3 we calculate the effective potential in the leading order of
the $1/N$ expansion up to linear terms in the space-time curvature.
In Sec.4 we evaluate the effective potential with varying the curvature
and the phase structure of the four-fermion theory is presented.
We derive the critical point where the phase transition is caused
by the curvature effect.
In Sec.5 our result is compared with the one obtained in
de Sitter space\cite{IMM} and Einstein universe\cite{IIM}
and we discuss the validity of the weak curvature expansion.

\subsection{Four-Fermion Theory in Curved Space-Time}

\hspace*{\parindent}
The four-fermion interaction theory is one of the prototype models of the
dynamical symmetry breaking.
In this paper we consider the simple four-fermion theory with $N$ components
of fermions\cite{GNRENG} described by the action
\begin{equation}
     S = \int \sqrt{-g} d^{\sD}x \left[
     \sum^{N}_{k=1}\bar{\psi}_{k}i\gamma_{\mu}\nabla^{\mu}\psi_{k}
     +\frac{\lambda_0}{2N}\sum^{N}_{k=1}(\bar{\psi}_{k}\psi_{k})^{2}
     \right]\, ,
\label{s:gn}
\end{equation}
where $g$ is the determinant of the space-time metric $g_{\mu\nu}$,
$D$ is the space-time dimension,
$\nabla^{\mu}\psi$ the covariant derivative of the fermion field $\psi$,
index $k$ represents the flavors of the fermion field
$\psi$ and $\lambda_0$ is a
bare coupling constant.  In the following, for simplicity,
we neglect the flavor index.
Our metric and curvature conventions are taken to be $(-,-,-)$
in the notation of the book by Misner, Thorne and Wheeler.\cite{MTW}
In two dimensions the theory is
nothing but the Gross-Neveu model.\cite{GN}
The theory has the discrete chiral symmetry,
$\bar\psi\psi \rightarrow -\bar\psi\psi$, and a global $SU(N)$ flavor symmetry.
The discrete chiral symmetry prevents the Lagrangian to have mass terms.
Under the circumstance of the global $SU(N)$ symmetry we may work in the
scheme of the $1/N$ expansion.

For practical calculations it is more convenient to introduce the
auxiliary field $\sigma$ and consider the following equivalent action
\begin{equation}
     S' = \int \sqrt{-g} d^{\sD}x
     \left[\bar{\psi}i\gamma_{\mu}\nabla^{\mu}\psi
     -\frac{N}{2\lambda_0}\sigma^{2}-\bar{\psi}\sigma\psi
     \right]\, .
\label{s:yukawa}
\end{equation}
Replacing $\sigma$ by the solution of the equation
of motion $\displaystyle \sigma \sim -\frac{\lambda_{0}}{N}\bar{\psi}\psi$
the action (\ref{s:gn}) is reproduced.
If the non-vanishing vacuum expectation value is assigned to
the auxiliary field $\sigma$, then there appears a mass term
for the fermion field $\psi$ and the discrete chiral
symmetry is eventually broken.

We would like to find a ground state of the system described
by the four-fermion theory.  For this purpose we evaluate an
effective potential for the auxiliary field $\sigma$.
To calculate the effective potential we start with the generating functional
$W[J]$ given by
\begin{equation}
     \exp (i N W[J])\equiv
     \int [d\sigma][d\psi][d\bar{\psi}]\exp i \left (S'
     +\int \sqrt{-g} d^{\sD}x\ \sigma J \right)\, ,
\label{wj}
\end{equation}
where $J$ is a source function.
We pull out an obvious factor $N$ in defining the generating functional
$W[J]$.
The effective action $\Gamma [\sigma]$
is defined to be the Legendre transform of
$W[J]$,
\begin{equation}
     \Gamma [\sigma]\equiv W[J]-\int \sqrt{-g} d^{\sD}x\ \sigma J\, .
\end{equation}
In the leading order of the $1/N$ expansion the effective action
$\Gamma [\sigma]$ is just equal to the semi-classical effective action.
Integrating over the fermion field in the Eq.(\ref{wj}) we obtain the
effective action of the theory considered here,
\begin{equation}
     \Gamma [\sigma]
     =-\int \sqrt{-g} d^{\sD}x\left[ \frac{1}{2\lambda_0}\sigma^{2}
                  +i\, \mbox{tr ln}(i \gamma_{\mu}\nabla^{\mu}-\sigma)
                  \right]+\mbox{O}(1/N)\, .
\label{eq:effac}
\end{equation}
{}From the effective action (\ref{eq:effac})
we obtain the effective potential $V(\sigma)$ in the
leading order of the $1/N$ expansion :
\begin{equation}
     V(\sigma ) = \frac{1}{2\lambda_0}\sigma^{2}
                  +i\, \mbox{tr ln}\frac{i \gamma_{\mu}\nabla^{\mu}-\sigma}
                                        {i \gamma_{\mu}\nabla^{\mu}}
                  +\mbox{O}(1/N)\, ,
\label{v:gn}
\end{equation}
where we normalize the effective potential so that $V(0)=0$.
In Eq.(\ref{v:gn}) the variable $\sigma$ is regarded as constant.

Using the Schwinger proper time method\cite{SP}
the second term on the right-hand side of Eq.(\ref{v:gn})
is rewritten as
\begin{equation}
      V(\sigma ) = \frac{1}{2\lambda_0}\sigma^{2}
                  -i\tr\int^{\sigma}_{0}ds\ S(x,x;s)+\mbox{O}(1/N)\, ,
\label{def:vs}
\end{equation}
where $S(x,x;s)$ is the two-point function of a massive free fermion
and satisfies the Dirac equation in curved space-time :
\begin{equation}
      (i \gamma_{\mu}\nabla^{\mu}-s)S(x,y;s)
      =\frac{1}{\sqrt{-g}}\delta^{\sD}(x,y)\, .
\end{equation}
Therefore the effective potential
is described by the two-point function $S(x,x;s)$ of the massive
free fermion in curved space-time.

\subsection{Effective Potential in Weak Curvature Expansion}

\hspace*{\parindent}
We would like to investigate the phase transition induced by curvature
effects in general curved space-time.
In this section we assume that the space-time curves slowly and neglect
terms involving higher derivatives of the metric tensor than the third
derivative.
As we have seen in the previous section,
the effective potential is obtained directly from the two-point
function $S(x,x;s)$ of a massive free fermion in curved space-time.
Thus we start with the argument on the two-point function $S(x,y;s)$
in the weak curvature expansion.

According to the method developed by Parker and Toms,\cite{PT}
the two-point function $S(x,x;s)$ is expanded asymptotically about $R=0$.
For this purpose
we introduce the bispinor function $G(x,y;s)$ defined by
\begin{equation}
     (i \gamma_{\mu}\nabla^{\mu}+s)G(x,y;s)=S(x,x;s)\, .
\label{def:g}
\end{equation}
It satisfies the following equation :
\begin{equation}
     \left(-\nabla^{\mu}\nabla_{\mu}-\frac{R}{4}-s^{2}\right)G(x,y;s)
     =\frac{1}{\sqrt{-g}}\delta^{\sD}(x,y)\, .
\label{eq:g}
\end{equation}

Using the Riemann normal coordinate\cite{RNC} with the origin at $x$
we expand the first term on the left-hand
side of Eq.(\ref{eq:g}) and find
\begin{equation}
\begin{array}{rcl}
     \displaystyle \sqrt{-g}\nabla^{\mu}\nabla_{\mu}G(x,y;s)&=& \displaystyle
     \left[\eta^{\mu\nu}\partial_{\mu}\partial_{\nu}
     +\frac{1}{6}{R^{0}}_{\alpha\beta}(y-x)^{\alpha}(y-x)^{\beta}
     \eta^{\mu\nu}\partial_{\mu}\partial_{\nu}\right.
\\[5mm]
     && \displaystyle -\frac{1}{3}{{{{R^{0}}^{\mu}}_{\alpha}}^{\nu}}_{\beta}
        (y-x)^{\alpha}(y-x)^{\beta}\partial_{\mu}\partial_{\nu}
        -\frac{2}{3}{{R^{0}}^{\mu}}_{\alpha}(y-x)^{\alpha}\partial_{\mu}
\\[5mm]
    && \displaystyle \left. +\frac{1}{4}{{R^{0}}^{\mu}}_{\alpha a b}
       \sigma^{ab}(y-x)^{\alpha}\partial_{\mu}+\cdots
     \right]G(x,y;s)\, ,
\end{array}
\label{eq:rne}
\end{equation}
where
\begin{equation}
     \sigma^{ab}=\frac{1}{2}[\gamma^{a},\gamma^{b}]\, ,
\end{equation}
and Latin indices $a,b$ are vierbein indices,
$\eta^{\mu\nu}$ the Minkowski metric and the suffix $0$
denotes tensors at the origin $x$.
Here we keep only terms independent of the curvature $R$ and terms linear
in $R$.
Inserting the Eq.(\ref{eq:rne}) into Eq.(\ref{eq:g}) and
performing the following Fourier transformation :
\begin{equation}
     G(x,y)=\int \frac{d^{\sD}p}{(2\pi)^{\sD}}e^{-ip(x-y)}\tilde{G}(p,y)\, ,
\label{fou:g}
\end{equation}
 Equation (\ref{eq:g}) reads
\begin{equation}
\begin{array}{l}
     \displaystyle \left[\eta^{\mu\nu}p_{\mu}p_{\nu}-\frac{1}{4}R^{0}-m^{2}
     -\frac{2}{3}{{R^{0}}^{\mu}}_{\alpha}p_{\mu}
     \frac{\partial}{\partial p_{\alpha}}
     -\frac{1}{6}{R^{0}}^{\alpha\beta}\eta^{\mu\nu}p_{\mu}p_{\nu}
     \frac{\partial}{\partial p_{\alpha}}\frac{\partial}{\partial p_{\beta}}
     \right.
\\[5mm]
     \displaystyle \left.
     +\frac{1}{3}{{{{R^{0}}^{\mu}}_{\alpha}}^{\nu}}_{\beta}p_{\mu}p_{\nu}
     \frac{\partial}{\partial p_{\alpha}}\frac{\partial}{\partial p_{\beta}}
     +\frac{1}{6}{R^{0}}^{\alpha\beta}m^{2}
     \frac{\partial}{\partial p_{\alpha}}\frac{\partial}{\partial p_{\beta}}
     -\frac{1}{4}{{R^{0}}^{\mu}}_{\alpha a b}\sigma^{ab}p_{\mu}
     \frac{\partial}{\partial p_{\alpha}}
     \right]\tilde{G}(p,y)=1\, .
\end{array}
\label{eq:tildeg}
\end{equation}
{}From the Eq.(\ref{eq:tildeg}) $\tilde{G}(p,y)$ is found to be
\begin{equation}
     \tilde{G}(p,y)=\frac{1}{p^{2}-m^{2}}
     -\frac{1}{12}\frac{R^{0}}{(p^{2}-m^{2})^{2}}
     +\frac{2}{3}\frac{{R^{0}}^{\mu\nu}p_{\mu}p_{\nu}}{(p^{2}-m^{2})^{3}}
     +\mbox{O}(R_{;\mu},R^{2})\, .
\label{exp:tildeg}
\end{equation}
Inserting the Eqs.(\ref{fou:g}) and (\ref{exp:tildeg}) into Eq.(\ref{def:g})
the spinor two-point function $S(x,y;s)$ in the weak curvature
expansion is obtained
\begin{equation}
\begin{array}{rcl}
     \displaystyle S(x,y;s)&=&\displaystyle (i \gamma_{\mu}\nabla^{\mu}+s)
     \int \frac{d^{\sD}p}{(2\pi)^{\sD}}e^{-ip(x-y)}\tilde{G}(p,y)
\\[5mm]
     &=&\displaystyle \int \frac{d^{\sD}p}{(2\pi)^{\sD}}e^{-ip(x-y)}\left[
        \frac{\gamma^{a}p_{a}+s}{p^{2}-s^{2}}
        -\frac{1}{12}R^{0}\frac{\gamma^{a}p_{a}+s}{(p^{2}-s^{2})^{2}}
     \right.
\\[5mm]
     &&\displaystyle\left.+\frac{2}{3}{R^{0}}^{\mu\nu}p_{\mu}p_{\nu}
     \frac{\gamma^{a}p_{a}+s}{(p^{2}-s^{2})^{3}}
     +\frac{1}{4}\gamma^{a}\sigma^{cd}{R^{0}}_{cda\mu}p^{\mu}
     \frac{1}{(p^{2}-s^{2})^{2}}
     \right]+\mbox{O}(R_{;\mu},R^{2})\, .
\end{array}
\label{exp:s}
\end{equation}

By substituting Eq.(\ref{exp:s}) into Eq.(\ref{def:vs})
and integrating over $p$ and $s$, we obtain
the effective potential $V(\sigma)$ up to linear terms in
the space-time curvature,
\begin{equation}
     V(\sigma)=V_{0}(\sigma)+V_{\sR}(\sigma)+\mbox{O}(R_{;\mu},R^{2})\, ,
\label{v:nonren}
\end{equation}
where $V_{0}(\sigma)$ is the effective potential at $R=0$,
\begin{equation}
     V_{0}(\sigma) = \frac{1}{2\lambda_0}\sigma^{2}
                 -\frac{\tr\11}{(4\pi)^{\sD/2}D}
                  \Gamma \left( 1-\frac{D}{2} \right)\sigma^{\sD} \, ,
\label{v0:nonren}
\end{equation}
and $V_{\sR}(\sigma)$ is the effective potential linear in $R$,
\begin{equation}
     V_{\sR}(\sigma)=-\frac{\tr \11}{(4\pi)^{\sD/2}}
     \frac{R}{24}\Gamma\left(1-\frac{D}{2}\right)\sigma^{\sD-2}\, ,
\label{vr:nonren}
\end{equation}
where $\tr \11$ represents the trace of the unit Dirac matrix.

We clearly see that the effective potential (\ref{v:nonren}) is
divergent in two and four dimensions. It happens to be
finite in three dimensions in the leading order of the $1/N$
expansion.
As is well-known, four-fermion theory is renormalizable in
two dimensional Minkowski space.  Therefore the potential
(\ref{v0:nonren}) is made finite at $D=2$ by the usual
renormalization procedure. In four
dimensions four-fermion theory is not renormalizable and the
finite effective potential can not be defined.
We regard the effective potential for
$D=4-\epsilon$ with $\epsilon$ sufficiently small positive
as a regularization of the one in four dimensions.

We perform the renormalization in two dimensions by imposing
the renormalization condition,
\begin{equation}
     \left.
     \frac{\partial^{2}V_{0}(\sigma)}{\partial \sigma^{2}}
     \right|_{\sigma = \mu}
     =\frac{\mu^{\sD-2}}{\lambda}\, ,
\label{cond:ren}
\end{equation}
where $\mu$ is the renormalization scale.  From
this renormalization condition we get the renormalized
coupling $\lambda$,
\begin{equation}
     \frac{1}{\lambda_0}=\frac{\mu^{\sD-2}}{\lambda}
                       +\frac{\tr\11}{(4\pi)^{\sD/2}}(D-1)
                        \Gamma \left( 1-\frac{D}{2} \right)
                        \mu^{\sD-2}\, .
\label{eqn:ren}
\end{equation}
Replacing the bare coupling constant $\lambda_{0}$ with the renormalized
one $\lambda$, we obtain the renormalized effective potential
\begin{equation}
\begin{array}{rcl}
     \displaystyle V(\sigma)
     & = &  \displaystyle \frac{1}{2\lambda}\sigma^{2}\mu^{\sD-2}
            +\frac{\tr \11}{2(4\pi)^{\sD/2}}(D-1)
             \Gamma \left( 1-\frac{D}{2}\right)
             \sigma^{2}\mu^{\sD-2}
\\[5mm]
     &   &  \displaystyle -\frac{\tr \11}{(4\pi)^{\sD/2}D}
             \Gamma \left( 1-\frac{D}{2} \right)\sigma^{\sD}
            -\frac{\tr \11}{(4\pi)^{\sD/2}}
            \frac{R}{24}\Gamma\left(1-\frac{D}{2}\right)\sigma^{\sD-2}
\, .
\end{array}
\label{v:ren}
\end{equation}
In Minkowski space, the renormalized effective potential,
$V(\sigma)|_{R=0}=V_{0}(\sigma)$, is no longer divergent in
the whole range of the
space-time dimensions considered here, $2 \leq D < 4$.
The effective potential (\ref{v:ren}) agrees with the weak curvature
limit of the results obtained in de Sitter space\cite{IMM} and Einstein
Universe.\cite{IIM}

Next we consider the two-, three- and four-dimensional limits of the
effective potential (\ref{v:ren}).
Taking the two dimensional limit, $D\rightarrow 2$, we get
\begin{equation}
\begin{array}{rcl}
     \displaystyle \frac{V(\sigma)}{\mu^{\sD}}&=& \displaystyle
                   \frac{1}{2 \lambda}\left(\frac{\sigma}{\mu}\right)^{2}
                  +\frac{\tr\11}{8 \pi}\left[-3+\ln\left(\frac{\sigma}{\mu}
                   \right)^{2}
                   \right]\left(\frac{\sigma}{\mu}\right)^{2}
\\[5mm]
     \displaystyle & &  \displaystyle -\frac{\tr\11}{96 \pi}\frac{R}{\mu^{2}}
      -\frac{\tr\11}{96 \pi}\frac{R}{\mu^{2}}
                        \left[\frac{2}{2-D}-\gamma+\ln 4\pi
                        -\ln \left(\frac{\sigma}{\mu}\right)^{2}
                        \right]\, .
\end{array}
\label{v:2d}
\end{equation}
It is different from the expression obtained in Ref.6.
The third term on the right-hand side in the Eq.(\ref{v:2d})
develops an infrared divergence.
The divergence appears from the mass singularity at
$\sigma\rightarrow 0$ and the normalization condition $V(0)=0$.
It causes the symmetry restoration for any positive values
of the space-time curvature $R$.

Taking the three dimensional limit $D\rightarrow 3$, we find
\begin{equation}
     \frac{V(\sigma)}{\mu^{3}}=
                   \frac{1}{2 \lambda}\left(\frac{\sigma}{\mu}\right)^{2}
                  -\frac{\tr\11}{4\pi}
                   \left[\left(\frac{\sigma}{\mu}\right)^{2}
                  -\frac{1}{3}\left(\frac{\sigma}{\mu}\right)^{3}\right]
     +\frac{\tr\11}{96 \pi}\frac{R}{\mu^{2}}
                        \frac{\sigma}{\mu}\, .
\label{v:3d}
\end{equation}
The effective potential (\ref{v:3d}) exactly reproduces the result obtained
in Ref.7.

If we take the four dimensional limit $D\rightarrow 4$, the effective
potential (\ref{v:ren}) reads
\begin{equation}
\begin{array}{rcl}
     \displaystyle \frac{V(\sigma)}{\mu^{\sD}}&=& \displaystyle
                   \frac{1}{2 \lambda}\left(\frac{\sigma}{\mu}\right)^{2}
     -\frac{\tr \11}{4(4 \pi)^{2}}
                   \left\{6\left(C-\frac{2}{3}\right)
                   \left(\frac{\sigma}{\mu}\right)^{2}
                   -\left[C+\frac{1}{2}
                   -\ln\left(\frac{\sigma}{\mu}\right)^{2}
                   \right]\left(\frac{\sigma}{\mu}\right)^{4}\right\}
\\[5mm]
     \displaystyle & &  \displaystyle
                   +\frac{\tr \11}{4(4 \pi)^{2}}\frac{1}{6}\frac{R}{\mu^{2}}
                        \left[C-\ln\left(\frac{\sigma}{\mu}\right)^{2}
                        \right]
                        \left(\frac{\sigma}{\mu}\right)^{2}\, ,
\end{array}
\label{v:4d}
\end{equation}
where we express the divergent parts as
\begin{equation}
     C=\frac{2}{4-D}-\gamma+\ln 4\pi +1\, .
\label{div:c}
\end{equation}
We find that there is a correspondence between this result
(\ref{v:4d}) and the result given in Ref.4 if we make a replacement
\begin{equation}
     C+\frac{R}{6}\frac{1}{6\mu^{2}-\sigma^{2}}
     \leftrightarrow \ln \frac{\Lambda^{2}}{\mu^{2}}\, ,
\end{equation}
where $\Lambda$ is a cut-off parameter of the divergent integral appearing
in Ref.4.
Note here that the direct comparison of our result
with the result in Ref.4 is possible only after renormalizing
the coupling constant $\lambda$ in Ref.4 under the renormalization
condition (\ref{cond:ren}).

\subsection{Phase Structure in Curved Space-Time}

\hspace*{\parindent}
The vacuum expectation value of the auxiliary field $\langle\sigma\rangle$
is determined by observing the minimum of the effective potential.
As is well-known in Minkowski space,\cite{IKM,GNRENG}
the minimum of the effective potential is located at the non-vanishing $\sigma$
for $\lambda > \lambda_{cr}$ and the chiral symmetry is broken down
dynamically.
A critical value of the coupling constant which divides the symmetric
and asymmetric phases is given by
\begin{equation}
     \lambda_{cr} = \frac{(4\pi)^{\sD/2}}{\tr\11}
                        \left[
                        (1-D)\Gamma \left( 1-\frac{D}{2} \right)
                        \right]^{-1}\, .
\label{cr:l:d}
\end{equation}

For $\lambda > \lambda_{cr}$ dynamical fermion mass is
generated.
It is obtained by the vacuum expectation value of
the auxiliary field $\langle\sigma\rangle$.
In the case of $R=0$ the dynamical fermion mass $m_{0}$ is known to be
\begin{equation}
     m_{0} = \langle\sigma\rangle = \mu
        \left[
                \frac{(4\pi)^{\sD/2}}
                        {\tr\11\Gamma \left( 1-D/2 \right)}
                \left(
                \frac{1}{\lambda}-\frac{1}{\lambda_{cr}}
                \right)
          \right]^{1/(\sD-2)} \, ,
\label{mass:d}
\end{equation}
where the suffix $0$ for $m_{0}$ is introduced to keep the memory that $R=0$.

\subsubsection{Phase Structure for $\lambda > \lambda_{cr}$}

\hspace*{\parindent}First we fix the coupling constant $\lambda$
no less than the critical value $\lambda_{cr}$ and see whether the
chiral symmetry is restored by curvature effects.
To study the phase structure in curved space-time we evaluate the effective
potential (\ref{v:ren}) with varying the space-time curvature.
In Fig.1 we present the typical behavior of the effective potential
(\ref{v:ren}) for several values of the curvature in the case of $D=2.5$
and $D=3.5$.
\begin{figure}
\setlength{\unitlength}{0.240900pt}
\begin{picture}(1500,900)(0,0)
\tenrm
\thicklines \path(220,113)(240,113)
\thicklines \path(1436,113)(1416,113)
\put(198,113){\makebox(0,0)[r]{-0.08}}
\thicklines \path(220,226)(240,226)
\thicklines \path(1436,226)(1416,226)
\put(198,226){\makebox(0,0)[r]{-0.06}}
\thicklines \path(220,339)(240,339)
\thicklines \path(1436,339)(1416,339)
\put(198,339){\makebox(0,0)[r]{-0.04}}
\thicklines \path(220,453)(240,453)
\thicklines \path(1436,453)(1416,453)
\put(198,453){\makebox(0,0)[r]{-0.02}}
\thicklines \path(220,566)(240,566)
\thicklines \path(1436,566)(1416,566)
\put(198,566){\makebox(0,0)[r]{0}}
\thicklines \path(220,679)(240,679)
\thicklines \path(1436,679)(1416,679)
\put(198,679){\makebox(0,0)[r]{0.02}}
\thicklines \path(220,792)(240,792)
\thicklines \path(1436,792)(1416,792)
\put(198,792){\makebox(0,0)[r]{0.04}}
\thicklines \path(220,113)(220,133)
\thicklines \path(220,877)(220,857)
\put(220,68){\makebox(0,0){0}}
\thicklines \path(363,113)(363,133)
\thicklines \path(363,877)(363,857)
\put(363,68){\makebox(0,0){0.2}}
\thicklines \path(506,113)(506,133)
\thicklines \path(506,877)(506,857)
\put(506,68){\makebox(0,0){0.4}}
\thicklines \path(649,113)(649,133)
\thicklines \path(649,877)(649,857)
\put(649,68){\makebox(0,0){0.6}}
\thicklines \path(792,113)(792,133)
\thicklines \path(792,877)(792,857)
\put(792,68){\makebox(0,0){0.8}}
\thicklines \path(935,113)(935,133)
\thicklines \path(935,877)(935,857)
\put(935,68){\makebox(0,0){1}}
\thicklines \path(1078,113)(1078,133)
\thicklines \path(1078,877)(1078,857)
\put(1078,68){\makebox(0,0){1.2}}
\thicklines \path(1221,113)(1221,133)
\thicklines \path(1221,877)(1221,857)
\put(1221,68){\makebox(0,0){1.4}}
\thicklines \path(1364,113)(1364,133)
\thicklines \path(1364,877)(1364,857)
\put(1364,68){\makebox(0,0){1.6}}
\thicklines \path(220,113)(1436,113)(1436,877)(220,877)(220,113)
\put(45,945){\makebox(0,0)[l]{\shortstack{$V/m_{0}^{2.5}$}}}
\put(828,23){\makebox(0,0){$\sigma/m_{0}$}}
\put(721,764){\makebox(0,0)[l]{$R=3R_{cr}/2$}}
\put(792,622){\makebox(0,0)[l]{$R=R_{cr}$}}
\put(792,481){\makebox(0,0)[l]{$R=R_{cr}/2$}}
\put(864,339){\makebox(0,0)[l]{$R=0$}}
\put(849,209){\makebox(0,0)[l]{$R=-R_{cr}/2$}}
\thinlines
\path(220,566)(220,566)(222,559)(223,556)(226,552)(233,546)
(245,537)(271,522)(321,492)(372,460)(423,425)(473,389)(524,352)(575,315)
(625,280)(676,246)(727,216)(777,190)(828,169)(853,160)(879,153)(891,150)
(904,148)(917,146)(929,144)(942,143)(948,142)(955,142)(958,142)(959,142)
(961,142)(963,142)(964,142)(966,142)(967,142)(969,142)(971,142)(972,142)
(974,142)(975,142)(977,142)(980,142)(983,142)(986,142)(993,142)(999,143)
(1005,143)(1018,145)(1031,147)(1056,153)(1081,161)
\thinlines
\path(1081,161)(1107,172)(1132,184)(1183,218)(1233,261)(1284,316)
(1335,382)(1385,461)(1436,552)
\thinlines
\path(220,566)(220,566)(222,566)(224,566)(226,566)(227,566)
(229,566)(231,565)(235,565)(242,565)(250,564)(265,561)(280,558)(309,550)
(339,540)(399,513)(458,482)(518,449)(578,414)(637,381)(697,351)(756,325)
(786,314)(816,305)(846,297)(861,294)(876,292)(891,290)(905,288)(913,288)
(917,287)(920,287)(924,287)(926,287)(928,287)(930,287)(932,287)(933,287)
(935,287)(937,287)(939,287)(941,287)(943,287)(945,287)(946,287)(950,287)
(954,287)(958,288)(965,288)(973,289)(980,290)
\thinlines
\path(980,290)(995,292)(1010,295)(1025,298)(1055,308)(1084,320)
(1114,336)(1174,377)(1233,432)(1293,502)(1353,589)(1412,693)(1436,742)
\thinlines
\path(220,566)(220,566)(222,573)(224,576)(227,580)(231,583)
(235,586)(242,590)(250,593)(257,595)(265,597)(272,599)(280,600)(283,600)
(287,600)(289,601)(291,601)(293,601)(295,601)(296,601)(298,601)(300,601)
(302,601)(304,601)(306,601)(308,601)(309,601)(311,601)(313,601)(317,601)
(324,600)(332,599)(339,598)(354,596)(369,593)(399,585)(458,565)(518,541)
(578,516)(637,491)(697,468)(756,449)(786,442)(816,436)(831,433)(846,431)
(853,431)(861,430)(868,429)(876,429)(879,429)
\thinlines
\path(879,429)(883,429)(885,429)(887,429)(889,429)(891,429)
(892,429)(894,429)(896,429)(898,429)(900,429)(902,429)(905,429)(907,429)
(909,429)(913,429)(920,429)(928,430)(935,430)(950,432)(965,435)(995,441)
(1025,451)(1055,463)(1114,496)(1174,543)(1233,603)(1293,678)(1353,770)
(1412,877)
\thinlines
\path(220,566)(220,566)(222,580)(224,586)(227,595)(231,601)
(235,607)(250,622)(257,628)(265,633)(280,641)(295,647)(309,652)(317,653)
(324,655)(332,656)(339,657)(343,657)(347,658)(350,658)(354,658)(356,658)
(358,658)(360,658)(362,658)(363,658)(365,658)(367,658)(369,658)(371,658)
(373,658)(375,658)(376,658)(380,658)(384,658)(391,658)(399,657)(414,656)
(429,653)(458,648)(518,634)(578,617)(637,601)(697,585)(727,579)(756,574)
(771,571)(786,569)(801,568)(816,567)(824,566)
\thinlines
\path(824,566)(831,566)(835,566)(837,566)(838,566)(840,566)
(842,566)(844,566)(846,566)(848,566)(850,566)(851,566)(853,566)(855,566)
(857,566)(861,566)(865,566)(868,566)(876,567)(883,567)(891,568)(905,569)
(920,571)(935,574)(965,581)(995,591)(1025,603)(1055,618)(1114,657)
(1174,708)(1233,774)(1293,854)(1307,877)
\thinlines
\path(220,566)(220,566)(222,588)(224,597)(227,610)(231,619)
(235,627)(250,652)(265,669)(280,683)(295,694)(309,702)(339,716)(354,720)
(369,724)(384,727)(399,729)(406,730)(414,730)(417,731)(421,731)(425,731)
(429,731)(432,731)(434,731)(436,731)(438,731)(440,731)(442,731)(444,731)
(445,731)(447,731)(449,731)(451,731)(455,731)(458,731)(466,731)(473,731)
(488,730)(503,728)(518,727)(578,719)(637,710)(667,706)(697,703)(712,701)
(727,700)(742,699)(756,698)(764,698)(771,697)
\thinlines
\path(771,697)(775,697)(777,697)(779,697)(781,697)(783,697)
(784,697)(786,697)(788,697)(790,697)(792,697)(794,697)(796,697)(797,697)
(801,697)(805,697)(809,698)(816,698)(824,698)(831,699)(846,700)(861,702)
(876,704)(905,710)(935,718)(965,728)(995,740)(1055,773)(1114,817)
(1174,874)(1176,877)
\dottedline{14}(220,566)(1436,566)
\end{picture}
                \vglue 1ex
                \hspace*{15em}\mbox{$(a) D=2.5$}
                \vglue 7ex
\setlength{\unitlength}{0.240900pt}
\begin{picture}(1500,900)(0,0)
\tenrm
\dottedline{14}(220,566)(1436,566)
\thicklines \path(220,113)(240,113)
\thicklines \path(1436,113)(1416,113)
\put(198,113){\makebox(0,0)[r]{-0.08}}
\thicklines \path(220,226)(240,226)
\thicklines \path(1436,226)(1416,226)
\put(198,226){\makebox(0,0)[r]{-0.06}}
\thicklines \path(220,339)(240,339)
\thicklines \path(1436,339)(1416,339)
\put(198,339){\makebox(0,0)[r]{-0.04}}
\thicklines \path(220,453)(240,453)
\thicklines \path(1436,453)(1416,453)
\put(198,453){\makebox(0,0)[r]{-0.02}}
\thicklines \path(220,566)(240,566)
\thicklines \path(1436,566)(1416,566)
\put(198,566){\makebox(0,0)[r]{0}}
\thicklines \path(220,679)(240,679)
\thicklines \path(1436,679)(1416,679)
\put(198,679){\makebox(0,0)[r]{0.02}}
\thicklines \path(220,792)(240,792)
\thicklines \path(1436,792)(1416,792)
\put(198,792){\makebox(0,0)[r]{0.04}}
\thicklines \path(220,113)(220,133)
\thicklines \path(220,877)(220,857)
\put(220,68){\makebox(0,0){0}}
\thicklines \path(363,113)(363,133)
\thicklines \path(363,877)(363,857)
\put(363,68){\makebox(0,0){0.2}}
\thicklines \path(506,113)(506,133)
\thicklines \path(506,877)(506,857)
\put(506,68){\makebox(0,0){0.4}}
\thicklines \path(649,113)(649,133)
\thicklines \path(649,877)(649,857)
\put(649,68){\makebox(0,0){0.6}}
\thicklines \path(792,113)(792,133)
\thicklines \path(792,877)(792,857)
\put(792,68){\makebox(0,0){0.8}}
\thicklines \path(935,113)(935,133)
\thicklines \path(935,877)(935,857)
\put(935,68){\makebox(0,0){1}}
\thicklines \path(1078,113)(1078,133)
\thicklines \path(1078,877)(1078,857)
\put(1078,68){\makebox(0,0){1.2}}
\thicklines \path(1221,113)(1221,133)
\thicklines \path(1221,877)(1221,857)
\put(1221,68){\makebox(0,0){1.4}}
\thicklines \path(1364,113)(1364,133)
\thicklines \path(1364,877)(1364,857)
\put(1364,68){\makebox(0,0){1.6}}
\thicklines \path(220,113)(1436,113)(1436,877)(220,877)(220,113)
\put(45,945){\makebox(0,0)[l]{\shortstack{$V/m_{0}^{3.5}$}}}
\put(828,23){\makebox(0,0){$\sigma/m_{0}$}}
\put(435,679){\makebox(0,0)[l]{$R=3R_{cr}/2$}}
\put(635,611){\makebox(0,0)[l]{$R=R_{cr}$}}
\put(663,520){\makebox(0,0)[l]{$R=R_{cr}/2$}}
\put(864,396){\makebox(0,0)[l]{$R=0$}}
\put(900,238){\makebox(0,0)[l]{$R=-R_{cr}/2$}}
\thinlines
\path(220,566)(220,566)(222,566)(223,566)(226,566)(230,565)
(233,565)(239,565)(245,564)(258,562)(271,560)(296,554)(321,547)
(372,527)(423,502)(473,472)(524,439)(575,403)(625,366)(676,328)
(727,291)(777,256)(828,224)(879,197)(929,176)(955,169)(967,166)
(980,163)(993,161)(999,160)(1005,160)(1008,159)(1012,159)(1015,159)
(1018,159)(1020,159)(1021,159)(1023,159)(1024,159)(1026,158)(1028,158)
(1029,158)(1031,158)(1032,159)(1034,159)(1035,159)(1037,159)(1040,159)
(1043,159)(1050,159)(1056,160)
\thinlines
\path(1056,160)(1069,162)(1081,165)(1094,168)(1107,172)
(1132,183)(1157,197)(1183,215)(1233,262)(1284,327)(1335,411)(1385,515)
(1436,642)
\thinlines
\path(220,566)(220,566)(222,566)(224,566)(226,566)(227,566)
(229,566)(231,566)(235,566)(239,565)(242,565)(250,565)(265,564)(280,562)
(309,557)(339,551)(399,534)(458,511)(518,485)(578,456)(637,427)(697,398)
(756,372)(786,360)(816,350)(846,342)(861,339)(876,336)(891,334)(905,332)
(913,331)(917,331)(920,331)(924,331)(928,331)(930,331)(932,331)(933,331)
(935,331)(937,331)(939,331)(941,331)(943,331)(945,331)(946,331)(950,331)
(954,331)(958,331)(965,332)(973,333)(980,334)
\thinlines
\path(980,334)(995,337)(1010,340)(1025,345)(1055,357)(1084,373)
(1114,393)(1174,448)(1233,526)(1293,627)(1353,756)(1398,877)
\thinlines
\path(220,566)(220,566)(227,566)(235,566)(239,566)(242,566)
(244,566)(246,566)(248,566)(250,566)(252,566)(254,566)(255,566)(257,566)
(259,566)(261,566)(265,566)(268,566)(272,566)(280,566)(287,565)(295,565)
(309,564)(324,563)(339,562)(369,558)(399,553)(458,542)(518,527)(578,512)
(637,496)(697,483)(727,477)(756,473)(771,472)(779,471)(786,470)(794,470)
(797,470)(801,470)(805,469)(809,469)(810,469)(812,469)(814,469)(816,469)
(818,469)(820,469)(822,469)(824,469)(825,469)
\thinlines
\path(825,469)(827,469)(831,469)(835,469)(838,470)(846,470)
(853,471)(861,471)(876,473)(891,475)(905,479)(935,487)(965,498)(995,513)
(1025,531)(1055,554)(1114,611)(1174,689)(1233,789)(1275,877)
\thinlines
\path(220,566)(220,566)(280,569)(295,570)(309,571)(324,572)
(339,572)(347,573)(354,573)(362,573)(369,573)(373,573)(375,573)(376,573)
(378,573)(380,573)(382,573)(384,573)(386,573)(388,573)(390,573)(391,573)
(393,573)(395,573)(399,573)(403,573)(406,573)(414,573)(429,572)(444,572)
(458,572)(518,569)(548,568)(578,567)(593,566)(600,566)(607,566)(611,566)
(615,566)(619,566)(622,566)(624,566)(626,566)(628,566)(630,566)(632,566)
(634,566)(635,566)(637,566)(639,566)(641,566)
\thinlines
\path(641,566)(645,566)(648,566)(652,566)(660,566)(667,566)
(675,567)(682,567)(697,568)(712,569)(727,570)(756,575)(786,580)(816,588)
(846,598)(876,610)(935,643)(995,689)(1055,750)(1114,830)(1142,877)
\thinlines
\path(220,566)(220,566)(280,573)(339,583)(399,592)(458,602)
(518,611)(578,622)(637,635)(697,653)(756,676)(816,707)(876,747)(935,799)
(995,865)(1004,877)
\end{picture}
		 \vglue 1ex
		 \hspace*{15em}\mbox{$(b) D=3.5$}
		 \vglue 1ex
\caption{Behaviors of the effective potential are shown at $D=2.5$
	  and $D=3.5$ for fixed $\lambda$ $( > \lambda_{cr})$
	  with varying the curvature where
	  $R_{cr}=6(D-2)(D(4-D)/4)^{(4-\sD)/(\sD-2)}m_{0}^{2}>0$.}
\label{fig:pot25}
\end{figure}
\noindent
Here we adopt the formula $\tr \11=2^{\sD/2}$.
We find that the chiral symmetry is restored as $R$ is
increased with $\lambda$ fixed.
As can be seen in Fig.1, the phase transition induced by
curvature effects is of the first order.
On the other hand the vacuum expectation value of the auxiliary field
is increased for a negative curvature.
We observe the similar behavior of the effective potential
in the space-time dimensions, $2 < D < 4$.
As is shown in Fig.1 that the derivative of the effective potential
for $R\neq 0$ is divergent at $\sigma=0$
in the space-time dimensions $2 \leq D < 3$ while that vanishes
in $3 \leq D < 4$.
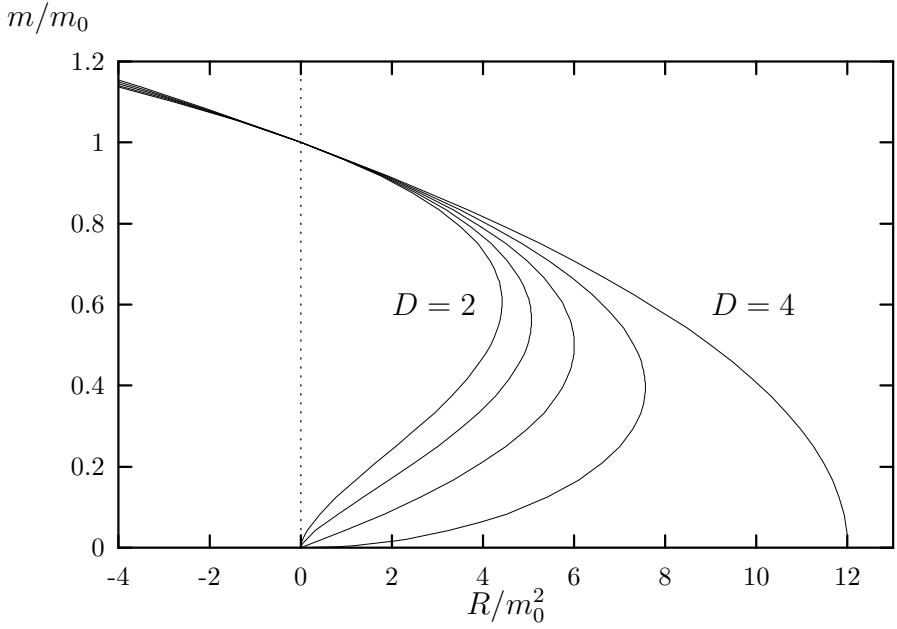
\begin{figure}
\setlength{\unitlength}{0.240900pt}
\begin{picture}(1500,900)(0,0)
\tenrm
\thinlines \dottedline{14}(506,113)(506,877)
\thicklines \path(220,113)(240,113)
\thicklines \path(1436,113)(1416,113)
\put(198,113){\makebox(0,0)[r]{0}}
\thicklines \path(220,240)(240,240)
\thicklines \path(1436,240)(1416,240)
\put(198,240){\makebox(0,0)[r]{0.2}}
\thicklines \path(220,368)(240,368)
\thicklines \path(1436,368)(1416,368)
\put(198,368){\makebox(0,0)[r]{0.4}}
\thicklines \path(220,495)(240,495)
\thicklines \path(1436,495)(1416,495)
\put(198,495){\makebox(0,0)[r]{0.6}}
\thicklines \path(220,622)(240,622)
\thicklines \path(1436,622)(1416,622)
\put(198,622){\makebox(0,0)[r]{0.8}}
\thicklines \path(220,750)(240,750)
\thicklines \path(1436,750)(1416,750)
\put(198,750){\makebox(0,0)[r]{1}}
\thicklines \path(220,877)(240,877)
\thicklines \path(1436,877)(1416,877)
\put(198,877){\makebox(0,0)[r]{1.2}}
\thicklines \path(220,113)(220,133)
\thicklines \path(220,877)(220,857)
\put(220,68){\makebox(0,0){-4}}
\thicklines \path(363,113)(363,133)
\thicklines \path(363,877)(363,857)
\put(363,68){\makebox(0,0){-2}}
\thicklines \path(506,113)(506,133)
\thicklines \path(506,877)(506,857)
\put(506,68){\makebox(0,0){0}}
\thicklines \path(649,113)(649,133)
\thicklines \path(649,877)(649,857)
\put(649,68){\makebox(0,0){2}}
\thicklines \path(792,113)(792,133)
\thicklines \path(792,877)(792,857)
\put(792,68){\makebox(0,0){4}}
\thicklines \path(935,113)(935,133)
\thicklines \path(935,877)(935,857)
\put(935,68){\makebox(0,0){6}}
\thicklines \path(1078,113)(1078,133)
\thicklines \path(1078,877)(1078,857)
\put(1078,68){\makebox(0,0){8}}
\thicklines \path(1221,113)(1221,133)
\thicklines \path(1221,877)(1221,857)
\put(1221,68){\makebox(0,0){10}}
\thicklines \path(1364,113)(1364,133)
\thicklines \path(1364,877)(1364,857)
\put(1364,68){\makebox(0,0){12}}
\thicklines \path(220,113)(1436,113)(1436,877)(220,877)(220,113)
\put(45,945){\makebox(0,0)[l]{\shortstack{$m / m_{0}$}}}
\put(828,23){\makebox(0,0){$R/m_{0}^2$}}
\put(1150,495){\makebox(0,0)[l]{$D=4$}}
\put(649,495){\makebox(0,0)[l]{$D=2$}}
\thinlines
\path(506,114)(506,114)(506,115)(506,116)(506,117)(507,118)(507,120)
(508,123)(509,126)(512,133)(516,140)(536,166)(562,193)(592,219)(623,246)
(655,272)(686,299)(716,325)(743,352)(767,378)(787,405)(796,418)(804,431)
(810,445)(815,458)(817,464)(819,471)(820,478)(821,484)(821,488)(821,489)
(822,491)(822,493)(822,494)(822,495)(822,496)(822,497)(822,498)(822,498)
(822,499)(822,500)(822,501)(822,502)(822,503)(822,504)(822,506)(822,508)
(821,511)(821,514)(820,518)(819,524)
\thinlines
\path(819,524)(818,531)(815,537)(810,551)(803,564)(784,591)
(757,617)(723,644)(682,670)(632,697)(573,723)(506,750)(506,750)(414,782)
(308,813)(220,837)
\thinlines
\path(506,113)(506,113)(506,114)(507,115)(507,116)(509,120)
(515,126)(529,140)(565,166)(604,193)(644,219)(684,246)(721,272)(755,299)
(785,325)(812,352)(833,378)(842,392)(850,405)(857,418)(862,431)(864,438)
(865,445)(867,451)(867,455)(867,458)(868,461)(868,463)(868,464)(868,466)
(868,467)(868,468)(868,469)(868,469)(868,470)(868,471)(868,472)(868,473)
(868,474)(868,474)(868,476)(868,478)(868,479)(868,481)(867,484)(867,488)
(867,491)(865,498)(864,504)(861,511)(856,524)
\thinlines
\path(856,524)(849,537)(830,564)(805,591)(773,617)(734,644)
(688,670)(634,697)(574,723)(506,750)(492,755)(398,787)(293,819)(220,839)
\thinlines
\path(506,113)(506,113)(575,140)(637,166)(694,193)(745,219)
(789,246)(828,272)(861,299)(888,325)(908,352)(917,365)(923,378)(929,392)
(931,398)(932,405)(934,411)(934,415)(935,418)(935,421)(935,423)(935,425)
(935,426)(935,427)(935,428)(935,429)(935,430)(935,431)(935,431)(935,432)
(935,433)(935,434)(935,435)(935,435)(935,436)(935,438)(935,440)(935,441)
(935,445)(934,448)(934,451)(932,458)(931,464)(929,471)(923,484)(917,498)
(908,511)(888,537)(861,564)(828,591)(789,617)
\thinlines
\path(789,617)(745,644)(694,670)(637,697)(575,723)(506,750)
(477,760)(382,793)(279,825)(220,842)
\thinlines
\path(506,113)(506,113)(547,114)(565,115)(589,116)(623,120)
(649,123)(671,126)(738,140)(788,153)(829,166)(893,193)(942,219)(979,246)
(1007,272)(1027,299)(1034,312)(1040,325)(1042,332)(1044,338)(1045,345)
(1046,352)(1046,355)(1047,358)(1047,360)(1047,361)(1047,362)(1047,363)
(1047,363)(1047,364)(1047,365)(1047,366)(1047,367)(1047,368)(1047,368)
(1047,369)(1047,370)(1047,372)(1047,373)(1046,375)(1046,378)(1046,382)
(1045,385)(1044,392)(1042,398)(1041,405)(1036,418)(1029,431)(1013,458)
(991,484)(964,511)
\thinlines
\path(964,511)(932,537)(895,564)(853,591)(807,617)(756,644)
(700,670)(640,697)(575,723)(506,750)(462,766)(368,798)(266,831)(220,845)
\thinlines
\path(1364,113)(1364,113)(1364,115)(1364,115)(1364,116)
(1364,117)(1364,118)(1364,120)(1364,121)(1364,123)(1364,126)(1364,130)
(1364,133)(1363,140)(1362,146)(1361,153)(1359,166)(1355,179)(1351,193)
(1341,219)(1327,246)(1311,272)(1291,299)(1269,325)(1244,352)(1215,378)
(1184,405)(1150,431)(1113,458)(1072,484)(1029,511)(983,537)(934,564)
(882,591)(827,617)(768,644)(707,670)(643,697)(576,723)(506,750)(448,771)
(354,804)(255,837)(220,848)
\end{picture}
\caption{Solutions of the gap equation for fixed $\lambda$ no less than
         $\lambda_{cr}$ at $D=2, 2.5, 3, 3.5, 4$.}
\label{fig:gap}
\end{figure}

In the leading order of the $1/N$ expansion
the dynamical fermion mass is equal to the vacuum expectation
value of the auxiliary field $\langle\sigma\rangle$.
We can find it by observing the minimum of the effective potential.
The necessary condition for the minimum is given by the gap equation :
\begin{equation}
      \left.\frac{\partial V(\sigma)}{\partial \sigma}\right|_{\sigma=m}
      =0\, .
\label{gap}
\end{equation}
The non-trivial solution of the gap equation corresponds to the
dynamical mass of the fermion.
Inserting the Eqs.(\ref{v:ren}) and (\ref{mass:d}) into Eq.(\ref{gap})
the non-trivial solution of the gap equation is expressed as
\begin{equation}
     m_{0}^{\sD-2}-m^{\sD-2}
     +\frac{R}{12}\left(1-\frac{D}{2}\right)m^{\sD-4}=0\, .
\label{nontri}
\end{equation}
In Fig.2 we plot the non-trivial solution of the gap equation
as a function of the space-time dimension $D$.

Since the phase transition induced by curvature effects
is of the first-order, two different solutions appear for
a small positive curvature.
The larger solution corresponds to the local minimum
and the smaller solution represents the first extremum
of the effective potential.

At the critical point the effective potential has the
same value at two local minimums.
Thus the critical point is obtained by the solution of Eq.(\ref{nontri})
which satisfies the condition
\begin{equation}
     V(m)=V(0)=0\, .
\label{cri1st}
\end{equation}
We solve the Eqs.(\ref{nontri}) and (\ref{cri1st})
and obtain the critical curvature,
\begin{equation}
     R_{cr}=6(D-2)\left(\frac{D(4-D)}{4}\right)^{(4-\sD)/(\sD-2)}m_{0}^{2}\, ,
\label{cr:r}
\end{equation}
and the mass gap at the critical point,
\begin{equation}
     m_{cr}=\left(\frac{D(4-D)}{4}\right)^{1/(\sD-2)}m_{0}\, .
\end{equation}

If the space-time curvature $R$ is no more than the critical value $R_{cr}$,
the vacuum expectation value of the auxiliary field $\langle\sigma\rangle$
is given by the larger solution of Eq.(\ref{nontri}).
If the space-time curvature $R$ is larger than the critical value $R_{cr}$,
the chiral symmetry is restored and the vacuum expectation value of the
auxiliary field $\langle\sigma\rangle$ disappears.
Thus the dynamical mass of the fermion looks in Fig.3.
\begin{figure}
\setlength{\unitlength}{0.240900pt}
\begin{picture}(1500,900)(0,0)
\tenrm
\thicklines \path(220,113)(240,113)
\thicklines \path(1436,113)(1416,113)
\put(198,113){\makebox(0,0)[r]{0}}
\thicklines \path(220,240)(240,240)
\thicklines \path(1436,240)(1416,240)
\put(198,240){\makebox(0,0)[r]{0.2}}
\thicklines \path(220,368)(240,368)
\thicklines \path(1436,368)(1416,368)
\put(198,368){\makebox(0,0)[r]{0.4}}
\thicklines \path(220,495)(240,495)
\thicklines \path(1436,495)(1416,495)
\put(198,495){\makebox(0,0)[r]{0.6}}
\thicklines \path(220,622)(240,622)
\thicklines \path(1436,622)(1416,622)
\put(198,622){\makebox(0,0)[r]{0.8}}
\thicklines \path(220,750)(240,750)
\thicklines \path(1436,750)(1416,750)
\put(198,750){\makebox(0,0)[r]{1}}
\thicklines \path(220,877)(240,877)
\thicklines \path(1436,877)(1416,877)
\put(198,877){\makebox(0,0)[r]{1.2}}
\thicklines \path(220,113)(220,133)
\thicklines \path(220,877)(220,857)
\put(220,68){\makebox(0,0){-4}}
\thicklines \path(363,113)(363,133)
\thicklines \path(363,877)(363,857)
\put(363,68){\makebox(0,0){-2}}
\thicklines \path(506,113)(506,133)
\thicklines \path(506,877)(506,857)
\put(506,68){\makebox(0,0){0}}
\thicklines \path(649,113)(649,133)
\thicklines \path(649,877)(649,857)
\put(649,68){\makebox(0,0){2}}
\thicklines \path(792,113)(792,133)
\thicklines \path(792,877)(792,857)
\put(792,68){\makebox(0,0){4}}
\thicklines \path(935,113)(935,133)
\thicklines \path(935,877)(935,857)
\put(935,68){\makebox(0,0){6}}
\thicklines \path(1078,113)(1078,133)
\thicklines \path(1078,877)(1078,857)
\put(1078,68){\makebox(0,0){8}}
\thicklines \path(1221,113)(1221,133)
\thicklines \path(1221,877)(1221,857)
\put(1221,68){\makebox(0,0){10}}
\thicklines \path(1364,113)(1364,133)
\thicklines \path(1364,877)(1364,857)
\put(1364,68){\makebox(0,0){12}}
\thicklines \path(220,113)(1436,113)(1436,877)(220,877)(220,113)
\put(45,945){\makebox(0,0)[l]{\shortstack{$m / m_{0}$}}}
\put(828,23){\makebox(0,0){$R/m_{0}^2$}}
\put(1150,495){\makebox(0,0)[l]{$D=4$}}
\put(1021,368){\makebox(0,0)[l]{$D=3.5$}}
\put(849,368){\makebox(0,0)[l]{$D=3$}}
\put(510,368){\makebox(0,0)[l]{$D=2.5$}}
\put(327,368){\makebox(0,0)[l]{$D=2$}}
\thinlines \path(506,750)(506,750)(506,750)(414,782)(308,813)(220,837)
\thinlines
\path(683,673)(683,673)(634,697)(574,723)(506,750)(492,755)
(398,787)(293,819)(220,839)
\thinlines
\path(828,591)(828,591)(789,617)(745,644)(694,670)(637,697)
(575,723)(506,750)(477,760)(382,793)(279,825)(220,842)
\thinlines
\path(995,480)(995,480)(991,484)(964,511)(932,537)(895,564)
(853,591)(807,617)(756,644)(700,670)(640,697)(575,723)(506,750)(462,766)
(368,798)(266,831)(220,845)
\thinlines
\path(1364,113)(1364,113)(1364,115)(1364,115)(1364,116)(1364,117)
(1364,118)(1364,120)(1364,121)(1364,123)(1364,126)(1364,130)(1364,133)
(1363,140)(1362,146)(1361,153)(1359,166)(1355,179)(1351,193)(1341,219)
(1327,246)(1311,272)(1291,299)(1269,325)(1244,352)(1215,378)(1184,405)
(1150,431)(1113,458)(1072,484)(1029,511)(983,537)(934,564)(882,591)
(827,617)(768,644)(707,670)(643,697)(576,723)(506,750)(448,771)(354,804)
(255,837)(220,848)
\dottedline{14}(506,750)(506,113)
\dottedline{14}(683,673)(683,113)
\dottedline{14}(828,591)(828,113)
\dottedline{14}(995,480)(995,113)
\end{picture}
\caption{Dynamical fermion mass as a function of the space-time curvature.}
\label{fig:mass}
\end{figure}
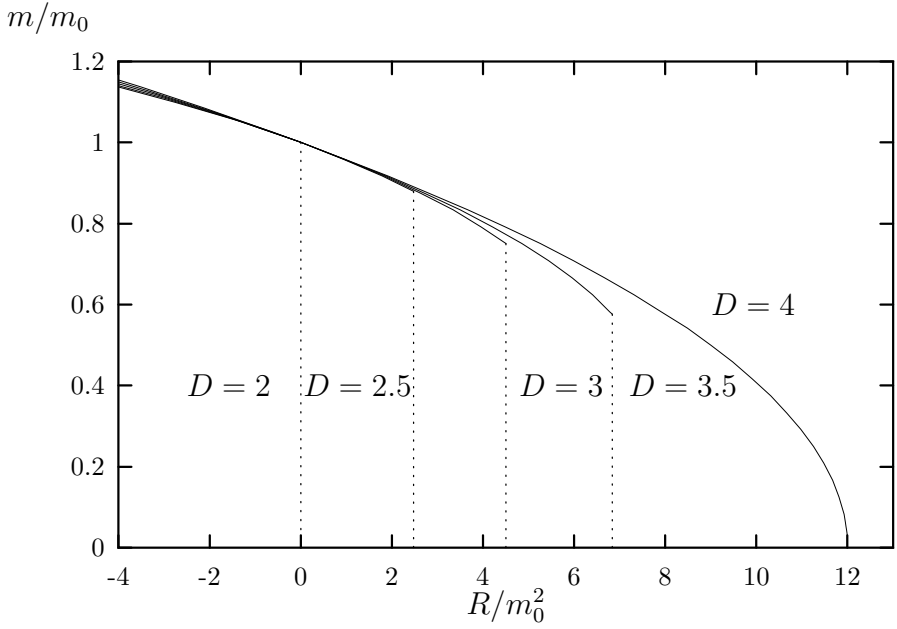
\noindent
We observe that the mass gap appears at $R=R_{cr}$ and disappears at the
four-dimensional limit.
For some special values of $D$ Eq.(\ref{cr:r}) simplifies :
\begin{equation}
\begin{array}{ll}
     R_{cr}=0\, ; & D=2\, ,
\\[5mm]
     \displaystyle R_{cr}=\frac{9}{2}m_{0}^{2}\, ; & D=3\, ,
\\[5mm]
     R_{cr}=12 m_{0}^{2}\, ; & D=4\, .
\end{array}
\label{rc234:wc}
\end{equation}

For some compact space-times we know the critical curvature without
making any approximation of the space-time curvature.
In the leading order of the $1/N$ expansion the exact expressions
of the critical curvature is given by
\begin{equation}
     R_{cr}=D(D-1)\left[\Gamma\left(\frac{D}{2}\right)
            \right]^{4/(2-\sD)}m_{0}^{2}\, ,
\label{rcr:desitter}
\end{equation}
in de Sitter space\cite{IMM} and
\begin{equation}
     R_{cr}=(D-1)(D-2)\left[
      \frac{1}{\sqrt{\pi}}
     \Gamma\left(\frac{D-1}{2}\right)\Gamma\left(\frac{D}{2}\right)
     \right]^{2/(2-\sD)}m_{0}^{2}\, ,
\label{rcr:einstein}
\end{equation}
in Einstein universe.\cite{IIM}

It is easy to see that the critical curvatures given
in Eqs.(\ref{rcr:desitter}) and (\ref{rcr:einstein})
reduce to
\begin{equation}
     R_{cr}=12 m_{0}^{2}\, ,
\end{equation}
at the four-dimensional limit.
Thus the weak curvature expansion gives the exact result for
$D=4$.
In Fig.4 we show the critical curvature $R_{cr}$ as a function of
the space-time dimensions $D$.
\begin{figure}
\setlength{\unitlength}{0.240900pt}
\begin{picture}(1500,900)(0,0)
\tenrm
\thicklines \path(220,113)(240,113)
\thicklines \path(1436,113)(1416,113)
\put(198,113){\makebox(0,0)[r]{0}}
\thicklines \path(220,235)(240,235)
\thicklines \path(1436,235)(1416,235)
\put(198,235){\makebox(0,0)[r]{2}}
\thicklines \path(220,357)(240,357)
\thicklines \path(1436,357)(1416,357)
\put(198,357){\makebox(0,0)[r]{4}}
\thicklines \path(220,480)(240,480)
\thicklines \path(1436,480)(1416,480)
\put(198,480){\makebox(0,0)[r]{6}}
\thicklines \path(220,602)(240,602)
\thicklines \path(1436,602)(1416,602)
\put(198,602){\makebox(0,0)[r]{8}}
\thicklines \path(220,724)(240,724)
\thicklines \path(1436,724)(1416,724)
\put(198,724){\makebox(0,0)[r]{10}}
\thicklines \path(220,846)(240,846)
\thicklines \path(1436,846)(1416,846)
\put(198,846){\makebox(0,0)[r]{12}}
\thicklines \path(220,113)(220,133)
\thicklines \path(220,877)(220,857)
\put(220,68){\makebox(0,0){2}}
\thicklines \path(342,113)(342,133)
\thicklines \path(342,877)(342,857)
\put(342,68){\makebox(0,0){2.2}}
\thicklines \path(463,113)(463,133)
\thicklines \path(463,877)(463,857)
\put(463,68){\makebox(0,0){2.4}}
\thicklines \path(585,113)(585,133)
\thicklines \path(585,877)(585,857)
\put(585,68){\makebox(0,0){2.6}}
\thicklines \path(706,113)(706,133)
\thicklines \path(706,877)(706,857)
\put(706,68){\makebox(0,0){2.8}}
\thicklines \path(828,113)(828,133)
\thicklines \path(828,877)(828,857)
\put(828,68){\makebox(0,0){3}}
\thicklines \path(950,113)(950,133)
\thicklines \path(950,877)(950,857)
\put(950,68){\makebox(0,0){3.2}}
\thicklines \path(1071,113)(1071,133)
\thicklines \path(1071,877)(1071,857)
\put(1071,68){\makebox(0,0){3.4}}
\thicklines \path(1193,113)(1193,133)
\thicklines \path(1193,877)(1193,857)
\put(1193,68){\makebox(0,0){3.6}}
\thicklines \path(1314,113)(1314,133)
\thicklines \path(1314,877)(1314,857)
\put(1314,68){\makebox(0,0){3.8}}
\thicklines \path(1436,113)(1436,133)
\thicklines \path(1436,877)(1436,857)
\put(1436,68){\makebox(0,0){4}}
\thicklines \path(220,113)(1436,113)(1436,877)(220,877)(220,113)
\put(45,945){\makebox(0,0)[l]{\shortstack{$R_{cr}/m_{0}^{2}$}}}
\put(828,23){\makebox(0,0){$D$}}
\put(342,755){\makebox(0,0)[l]{Symmetric Phase}}
\put(919,235){\makebox(0,0)[l]{Broken Phase}}
\thinlines
\path(220,113)(220,113)(221,113)(271,143)(321,170)(372,195)
(422,219)(473,242)(523,264)(573,285)(624,306)(674,326)(724,346)(775,366)
(825,387)(876,408)(926,429)(976,452)(1027,476)(1077,501)(1128,528)
(1178,558)(1228,592)(1279,630)(1329,676)(1354,704)(1380,735)(1405,774)
(1430,826)(1436,846)
\thicklines
\put(222,501){\circle*{0.4}}
\put(223,502){\circle*{0.4}}
\put(226,504){\circle*{0.4}}
\put(233,506){\circle*{0.4}}
\put(245,512){\circle*{0.4}}
\put(271,522){\circle*{0.4}}
\put(321,543){\circle*{0.4}}
\put(372,562){\circle*{0.4}}
\put(423,581){\circle*{0.4}}
\put(473,599){\circle*{0.4}}
\put(524,616){\circle*{0.4}}
\put(575,633){\circle*{0.4}}
\put(625,649){\circle*{0.4}}
\put(676,664){\circle*{0.4}}
\put(727,679){\circle*{0.4}}
\put(777,694){\circle*{0.4}}
\put(828,708){\circle*{0.4}}
\put(879,721){\circle*{0.4}}
\put(929,734){\circle*{0.4}}
\put(980,747){\circle*{0.4}}
\put(1031,759){\circle*{0.4}}
\put(1081,771){\circle*{0.4}}
\put(1132,783){\circle*{0.4}}
\put(1183,794){\circle*{0.4}}
\put(1233,805){\circle*{0.4}}
\put(1284,816){\circle*{0.4}}
\put(1335,826){\circle*{0.4}}
\put(1385,837){\circle*{0.4}}
\put(1436,846){\circle*{0.4}}
\thinlines
\dashline[3]{8}(222,115)(222,115)(223,117)(226,121)(233,129)
(245,144)(271,174)(321,230)(372,281)(423,327)(473,370)(524,410)(575,447)
(625,482)(676,515)(727,546)(777,575)(828,602)(879,628)(929,653)(980,676)
(1031,698)(1081,720)(1132,740)(1183,760)(1233,779)(1284,797)(1335,814)
(1385,830)(1436,846)
\end{picture}
\caption{Critical curvature as a function of dimension $D$.
The full line represents the critical curvature obtained by
the weak curvature expansion. The dotted and the dashed lines
represent the exact solutions in de Sitter space and Einstein
universe respectively.
}
\label{fig:crcurall}
\end{figure}
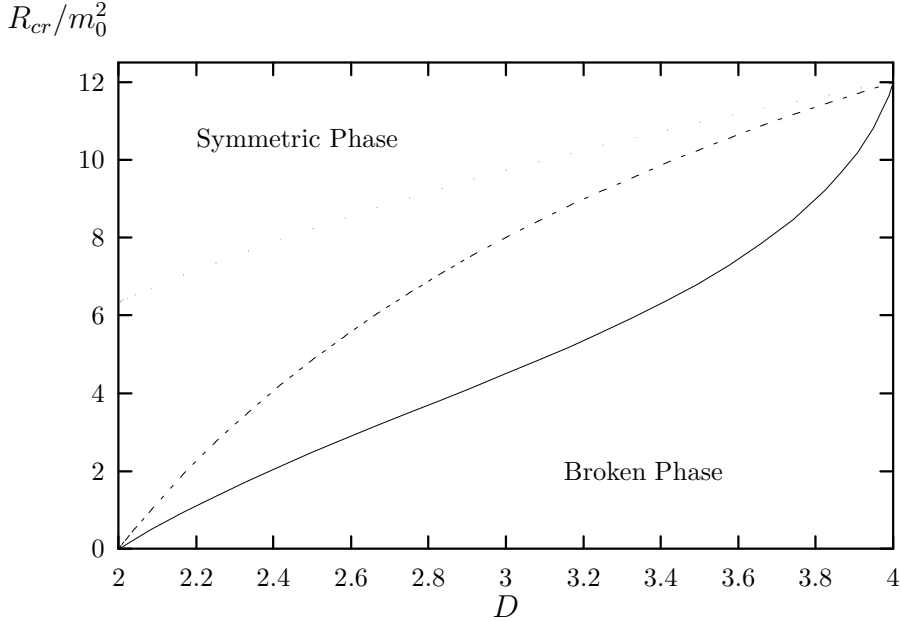
\noindent
It is clearly seen in Fig.4 that three lines of the critical curvature
reach the same value $12 m_{0}^{2}$ at $D\rightarrow 4$.
For $D=2$ the critical curvature obtained by the weak curvature expansion
is exactly equal to that in Einstein universe.
In the weak curvature expansion
the broken chiral symmetry is restored by the effect of an
infra-red divergence for any positive curvature at $D=2$.
The critical curvature is thus zero.
In two-dimensional Einstein universe the critical curvature
is also zero.
The situation is, however, different from the result by the
weak curvature expansion.
By definition two-dimensional Einstein universe is a flat space-time,
$R=0$.
The phase transition is induced by the topological effects
of the compact space.
The space-time curvature $R$ is not suitable to represent the phase structure
in two-dimensional Einstein universe.

\subsubsection{Phase Structure for $\lambda \leq \lambda_{cr}$}

\hspace*{\parindent}Next we fix the coupling constant $\lambda$ smaller than
the critical value
$\lambda_{cr}$ and study the phase structure in curved space-time.
\begin{figure}
\setlength{\unitlength}{0.240900pt}
\begin{picture}(1500,900)(0,0)
\tenrm
\thinlines \dottedline{14}(220,368)(1436,368)
\thicklines \path(220,113)(240,113)
\thicklines \path(1436,113)(1416,113)
\put(198,113){\makebox(0,0)[r]{-0.02}}
\thicklines \path(220,240)(240,240)
\thicklines \path(1436,240)(1416,240)
\put(198,240){\makebox(0,0)[r]{-0.01}}
\thicklines \path(220,368)(240,368)
\thicklines \path(1436,368)(1416,368)
\put(198,368){\makebox(0,0)[r]{0}}
\thicklines \path(220,495)(240,495)
\thicklines \path(1436,495)(1416,495)
\put(198,495){\makebox(0,0)[r]{0.01}}
\thicklines \path(220,622)(240,622)
\thicklines \path(1436,622)(1416,622)
\put(198,622){\makebox(0,0)[r]{0.02}}
\thicklines \path(220,750)(240,750)
\thicklines \path(1436,750)(1416,750)
\put(198,750){\makebox(0,0)[r]{0.03}}
\thicklines \path(220,877)(240,877)
\thicklines \path(1436,877)(1416,877)
\put(198,877){\makebox(0,0)[r]{0.04}}
\thicklines \path(220,113)(220,133)
\thicklines \path(220,877)(220,857)
\put(220,68){\makebox(0,0){0}}
\thicklines \path(423,113)(423,133)
\thicklines \path(423,877)(423,857)
\put(423,68){\makebox(0,0){0.05}}
\thicklines \path(625,113)(625,133)
\thicklines \path(625,877)(625,857)
\put(625,68){\makebox(0,0){0.1}}
\thicklines \path(828,113)(828,133)
\thicklines \path(828,877)(828,857)
\put(828,68){\makebox(0,0){0.15}}
\thicklines \path(1031,113)(1031,133)
\thicklines \path(1031,877)(1031,857)
\put(1031,68){\makebox(0,0){0.2}}
\thicklines \path(1233,113)(1233,133)
\thicklines \path(1233,877)(1233,857)
\put(1233,68){\makebox(0,0){0.25}}
\thicklines \path(1436,113)(1436,133)
\thicklines \path(1436,877)(1436,857)
\put(1436,68){\makebox(0,0){0.3}}
\thicklines \path(220,113)(1436,113)(1436,877)(220,877)(220,113)
\put(45,945){\makebox(0,0)[l]{\shortstack{$V/\sigma_{0}^{2.5}$}}}
\put(828,23){\makebox(0,0){$\sigma/\sigma_{0}$}}
\put(504,597){\makebox(0,0)[l]{$R=K/2$}}
\put(706,495){\makebox(0,0)[l]{$R=0$}}
\put(787,419){\makebox(0,0)[l]{$R=-K/2$}}
\put(1112,240){\makebox(0,0)[l]{$R=-K$}}
\thinlines
\path(220,368)(220,368)(222,368)(223,368)(225,368)(226,368)
(228,368)(230,368)(233,368)(236,368)(239,368)(245,368)(252,368)(258,368)
(271,368)(283,369)(296,369)(321,370)(347,371)(372,373)(423,377)(473,382)
(524,389)(575,397)(625,407)(676,418)(727,431)(777,445)(828,460)(879,477)
(929,496)(980,516)(1031,538)(1081,562)(1132,587)(1183,614)(1233,642)
(1284,672)(1335,704)(1385,738)(1436,774)
\thinlines
\path(220,368)(220,368)(222,374)(223,377)(226,380)(233,386)
(245,393)(271,404)(321,421)(372,435)(423,449)(473,463)(524,478)(575,493)
(625,509)(676,526)(727,545)(777,564)(828,585)(879,607)(929,631)(980,656)
(1031,683)(1081,711)(1132,740)(1183,771)(1233,804)(1284,838)(1335,874)
(1339,877)
\thinlines
\path(220,368)(220,368)(222,361)(223,359)(226,355)(233,350)
(239,346)(245,342)(271,332)(296,325)(321,319)(347,314)(372,310)(397,307)
(423,305)(435,304)(448,303)(461,302)(473,302)(486,301)(492,301)(499,301)
(502,301)(505,301)(508,301)(511,301)(513,301)(515,301)(516,301)(518,301)
(519,301)(521,301)(522,301)(524,301)(526,301)(527,301)(530,301)(533,301)
(537,301)(543,301)(549,301)(562,301)(575,302)(600,303)(625,305)(676,310)
(727,316)(777,325)(828,335)(879,347)(929,361)
\thinlines
\path(929,361)(980,376)(1031,394)(1081,413)(1132,433)(1183,456)
(1233,481)(1284,507)(1335,535)(1385,565)(1436,597)
\thinlines
\path(220,368)(220,368)(222,355)(223,350)(226,342)(233,332)
(245,317)(258,305)(271,296)(321,268)(372,248)(423,232)(473,221)(499,216)
(524,212)(549,209)(575,206)(600,204)(625,203)(638,202)(644,202)(651,202)
(657,202)(663,201)(666,201)(670,201)(673,201)(674,201)(676,201)(678,201)
(679,201)(681,201)(682,201)(684,201)(686,201)(687,201)(689,201)(690,201)
(692,201)(695,201)(701,201)(708,202)(714,202)(727,202)(739,203)(752,203)
(777,205)(803,207)(828,210)(879,217)(929,226)
\thinlines
\path(929,226)(980,236)(1031,249)(1081,264)(1132,280)(1183,299)
(1233,319)(1284,341)(1335,366)(1385,392)(1436,420)
\end{picture}
                \vglue 1ex
                \hspace*{15em}\mbox{$(a) D=2.5$}
                \vglue 7ex
\setlength{\unitlength}{0.240900pt}
\begin{picture}(1500,900)(0,0)
\tenrm
\thinlines \dottedline{14}(220,368)(1436,368)
\thicklines \path(220,113)(240,113)
\thicklines \path(1436,113)(1416,113)
\put(198,113){\makebox(0,0)[r]{-0.0015}}
\thicklines \path(220,198)(240,198)
\thicklines \path(1436,198)(1416,198)
\put(198,198){\makebox(0,0)[r]{-0.001}}
\thicklines \path(220,283)(240,283)
\thicklines \path(1436,283)(1416,283)
\put(198,283){\makebox(0,0)[r]{-0.0005}}
\thicklines \path(220,368)(240,368)
\thicklines \path(1436,368)(1416,368)
\put(198,368){\makebox(0,0)[r]{0}}
\thicklines \path(220,453)(240,453)
\thicklines \path(1436,453)(1416,453)
\put(198,453){\makebox(0,0)[r]{0.0005}}
\thicklines \path(220,537)(240,537)
\thicklines \path(1436,537)(1416,537)
\put(198,537){\makebox(0,0)[r]{0.001}}
\thicklines \path(220,622)(240,622)
\thicklines \path(1436,622)(1416,622)
\put(198,622){\makebox(0,0)[r]{0.0015}}
\thicklines \path(220,707)(240,707)
\thicklines \path(1436,707)(1416,707)
\put(198,707){\makebox(0,0)[r]{0.002}}
\thicklines \path(220,792)(240,792)
\thicklines \path(1436,792)(1416,792)
\put(198,792){\makebox(0,0)[r]{0.0025}}
\thicklines \path(220,877)(240,877)
\thicklines \path(1436,877)(1416,877)
\put(198,877){\makebox(0,0)[r]{0.003}}
\thicklines \path(220,113)(220,133)
\thicklines \path(220,877)(220,857)
\put(220,68){\makebox(0,0){0}}
\thicklines \path(423,113)(423,133)
\thicklines \path(423,877)(423,857)
\put(423,68){\makebox(0,0){0.05}}
\thicklines \path(625,113)(625,133)
\thicklines \path(625,877)(625,857)
\put(625,68){\makebox(0,0){0.1}}
\thicklines \path(828,113)(828,133)
\thicklines \path(828,877)(828,857)
\put(828,68){\makebox(0,0){0.15}}
\thicklines \path(1031,113)(1031,133)
\thicklines \path(1031,877)(1031,857)
\put(1031,68){\makebox(0,0){0.2}}
\thicklines \path(1233,113)(1233,133)
\thicklines \path(1233,877)(1233,857)
\put(1233,68){\makebox(0,0){0.25}}
\thicklines \path(1436,113)(1436,133)
\thicklines \path(1436,877)(1436,857)
\put(1436,68){\makebox(0,0){0.3}}
\thicklines \path(220,113)(1436,113)(1436,877)(220,877)(220,113)
\put(45,945){\makebox(0,0)[l]{\shortstack{$V/\sigma_{0}^{3.5}$}}}
\put(828,23){\makebox(0,0){$\sigma/\sigma_{0}$}}
\put(321,622){\makebox(0,0)[l]{$R=K/2$}}
\put(747,622){\makebox(0,0)[l]{$R=0$}}
\put(1071,622){\makebox(0,0)[l]{$R=-K/2$}}
\put(787,266){\makebox(0,0)[l]{$R=-K$}}
\thinlines
\path(220,368)(220,368)(222,368)(223,368)(225,368)(226,368)
(228,368)(230,368)(233,368)(236,368)(239,368)(245,368)(252,369)(258,369)
(271,370)(283,372)(296,373)(321,378)(347,384)(372,391)(423,409)(473,433)
(524,461)(575,496)(625,535)(676,580)(727,631)(777,688)(828,750)(879,819)
(919,877)
\thinlines
\path(220,368)(220,368)(222,368)(223,368)(226,368)(230,368)
(233,369)(245,371)(258,373)(271,377)(296,385)(321,396)(372,425)(423,461)
(473,506)(524,558)(575,617)(625,683)(676,757)(727,838)(749,877)
\thinlines
\path(220,368)(220,368)(271,364)(321,359)(334,359)(347,358)
(359,357)(372,357)(378,357)(382,357)(385,357)(388,357)(391,357)(393,356)
(394,356)(396,356)(397,356)(399,356)(400,356)(402,356)(404,356)(405,356)
(407,356)(410,356)(412,357)(413,357)(416,357)(423,357)(429,357)(435,357)
(448,358)(461,358)(473,359)(499,362)(524,365)(575,374)(625,387)(676,404)
(727,424)(777,449)(828,478)(879,512)(929,550)(980,593)(1031,641)(1081,693)
(1132,752)(1183,815)(1228,877)
\thinlines
\path(220,368)(220,368)(222,368)(223,367)(226,367)(233,366)
(245,364)(271,357)(321,341)(372,323)(423,304)(473,286)(524,269)(575,253)
(625,239)(676,227)(727,217)(752,213)(777,210)(790,209)(803,208)(815,207)
(828,206)(834,206)(841,205)(847,205)(853,205)(856,205)(860,205)(863,205)
(864,205)(866,205)(868,205)(869,205)(871,205)(872,205)(874,205)(876,205)
(877,205)(879,205)(880,205)(882,205)(885,205)(888,205)(891,205)(898,205)
(904,205)(917,206)(929,207)(955,209)(980,212)
\thinlines
\path(980,212)(1005,216)(1031,222)(1081,235)(1132,252)(1183,273)
(1233,298)(1284,329)(1335,364)(1385,403)(1436,448)
\end{picture}
                \vglue 1ex
                \hspace*{15em}\mbox{$(b) D=3.5$}
                \vglue 1ex
\caption{Behaviors of the effective potential are shown at $D=2.5$
         and $D=3.5$ for fixed $\lambda$ $( \leq \lambda_{cr})$
         with varying the curvature where
         $K=6(D-2)(D(4-D)/4)^{(4-\sD)/(\sD-2)}\sigma_{0}^{2}>0$.}
\label{fig:potsc}
\end{figure}
\noindent
We introduce, for convenience, a scale $\sigma_{0}$ defined by
\begin{equation}
\sigma_{0}= \mu
        \left[
                \frac{(4\pi)^{\sD/2}}
                        {\tr\11\Gamma \left( 1-D/2 \right)}
                \left(
                \frac{1}{\lambda_{cr}}-\frac{1}{\lambda}
                \right)
          \right]^{1/(\sD-2)} \, .
\end{equation}
In Fig.5 the behavior of the effective potential
(\ref{v:ren}) is presented with varying
the curvature at $D=2.5$ and $D=3.5$.
We observe only the second order phase transition as the curvature is
decreased.
The chiral symmetry is broken down for any negative values
of the curvature.
Thus the critical curvature is zero in the whole range of
the space-time dimensions $D$
considered here, $2 \leq D < 4$.
As is seen in Fig.5 that the derivative of the effective potential is
divergent at $\sigma=0$ for $R\neq 0$ in $2 < D < 3$.
In $3 < D < 4$ the derivative of the effective potential vanishes
at $\sigma=0$.

In the case $\lambda \leq \lambda_{cr}$ Eq.(\ref{nontri}) is modified as
\begin{equation}
     -\sigma_{0}^{\sD-2}-m^{\sD-2}
     +\frac{R}{12}\left(1-\frac{D}{2}\right)m^{\sD-4}=0\, .
\label{nontri2}
\end{equation}
The dynamical mass of the fermion is given by the solution of the
Eq.(\ref{nontri2}).
In Fig.6 we present the dynamical mass of the fermion as a function of
the space-time dimensions $D$.
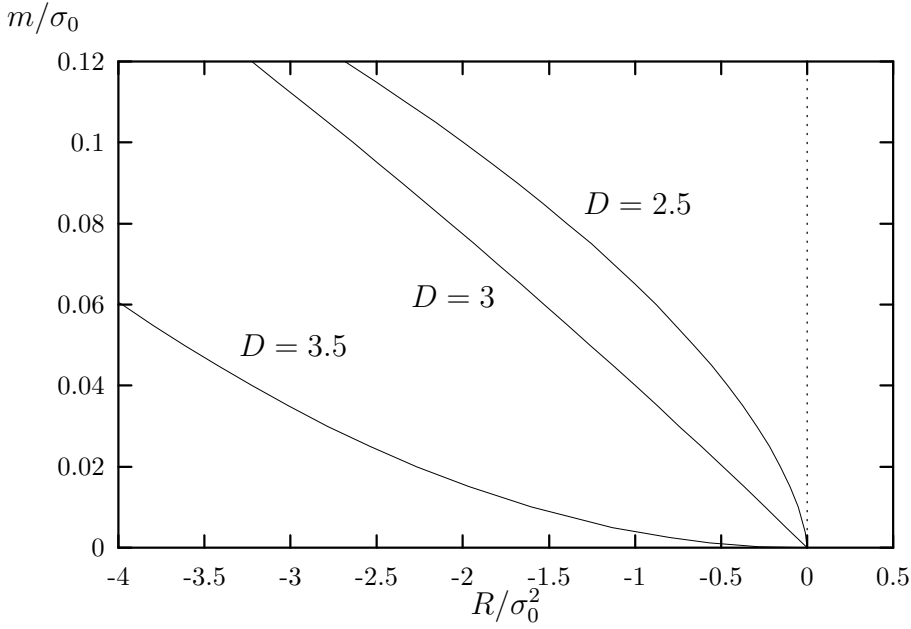
\begin{figure}
\setlength{\unitlength}{0.240900pt}
\begin{picture}(1500,900)(0,0)
\tenrm
\thinlines \dottedline{14}(1301,113)(1301,877)
\thicklines \path(220,113)(240,113)
\thicklines \path(1436,113)(1416,113)
\put(198,113){\makebox(0,0)[r]{0}}
\thicklines \path(220,240)(240,240)
\thicklines \path(1436,240)(1416,240)
\put(198,240){\makebox(0,0)[r]{0.02}}
\thicklines \path(220,368)(240,368)
\thicklines \path(1436,368)(1416,368)
\put(198,368){\makebox(0,0)[r]{0.04}}
\thicklines \path(220,495)(240,495)
\thicklines \path(1436,495)(1416,495)
\put(198,495){\makebox(0,0)[r]{0.06}}
\thicklines \path(220,622)(240,622)
\thicklines \path(1436,622)(1416,622)
\put(198,622){\makebox(0,0)[r]{0.08}}
\thicklines \path(220,750)(240,750)
\thicklines \path(1436,750)(1416,750)
\put(198,750){\makebox(0,0)[r]{0.1}}
\thicklines \path(220,877)(240,877)
\thicklines \path(1436,877)(1416,877)
\put(198,877){\makebox(0,0)[r]{0.12}}
\thicklines \path(220,113)(220,133)
\thicklines \path(220,877)(220,857)
\put(220,68){\makebox(0,0){-4}}
\thicklines \path(355,113)(355,133)
\thicklines \path(355,877)(355,857)
\put(355,68){\makebox(0,0){-3.5}}
\thicklines \path(490,113)(490,133)
\thicklines \path(490,877)(490,857)
\put(490,68){\makebox(0,0){-3}}
\thicklines \path(625,113)(625,133)
\thicklines \path(625,877)(625,857)
\put(625,68){\makebox(0,0){-2.5}}
\thicklines \path(760,113)(760,133)
\thicklines \path(760,877)(760,857)
\put(760,68){\makebox(0,0){-2}}
\thicklines \path(896,113)(896,133)
\thicklines \path(896,877)(896,857)
\put(896,68){\makebox(0,0){-1.5}}
\thicklines \path(1031,113)(1031,133)
\thicklines \path(1031,877)(1031,857)
\put(1031,68){\makebox(0,0){-1}}
\thicklines \path(1166,113)(1166,133)
\thicklines \path(1166,877)(1166,857)
\put(1166,68){\makebox(0,0){-0.5}}
\thicklines \path(1301,113)(1301,133)
\thicklines \path(1301,877)(1301,857)
\put(1301,68){\makebox(0,0){0}}
\thicklines \path(1436,113)(1436,133)
\thicklines \path(1436,877)(1436,857)
\put(1436,68){\makebox(0,0){0.5}}
\thicklines \path(220,113)(1436,113)(1436,877)(220,877)(220,113)
\put(45,945){\makebox(0,0)[l]{\shortstack{$m / \sigma_{0}$}}}
\put(828,23){\makebox(0,0){$R/\sigma_{0}^2$}}
\put(950,654){\makebox(0,0)[l]{$D=2.5$}}
\put(679,508){\makebox(0,0)[l]{$D=3$}}
\put(409,431){\makebox(0,0)[l]{$D=3.5$}}
\thinlines
\path(1301,113)(1301,113)(1301,114)(1301,115)(1301,117)
(1301,119)(1300,121)(1299,129)(1298,137)(1296,145)(1287,177)(1274,209)
(1259,240)(1242,272)(1222,304)(1200,336)(1176,368)(1151,400)(1123,431)
(1094,463)(1064,495)(1031,527)(997,559)(962,591)(924,622)(886,654)(846,686)
(804,718)(761,750)(717,782)(671,813)(624,845)(575,877)
\thinlines
\path(1301,113)(1301,113)(1268,145)(1235,177)(1202,209)
(1169,240)(1135,272)(1100,304)(1066,336)(1031,368)(996,400)(960,431)
(925,463)(888,495)(852,527)(815,559)(778,591)(741,622)(703,654)(665,686)
(626,718)(588,750)(548,782)(509,813)(469,845)(429,877)
\thinlines
\path(1301,113)(1301,113)(1247,114)(1224,115)(1193,117)
(1148,121)(1085,129)(995,145)(868,177)(770,209)(688,240)(615,272)
(548,304)(487,336)(429,368)(375,400)(323,431)(274,463)(226,495)(220,499)
\end{picture}
\caption{Solutions of the gap equation for fixed $\lambda$ smaller than
         $\lambda_{cr}$ at $D=2.5, 3, 3.5$.}
\label{fig:solsc}
\end{figure}
\noindent
As is expected for the second order phase transition,
the dynamical mass of the fermion smoothly disappears at $R=R_{cr}=0$.

\subsection{Discussions}

\hspace*{\parindent}
We expect that it will be one of the powerful approaches
for testing composite Higgs models at the GUT era
to study the dynamical symmetry breaking in curved space-time.
We have considered the Gross-Neveu type model as one of the prototype
models of the dynamical symmetry breaking and investigated the phase
structure in the weak curvature expansion.

In the leading order of the $1/N$ expansion the effective potential
is described by the two-point function of a massive free fermion.
Assuming that the space-time curvature is small we calculated
the two-point function up to linear terms in the curvature and obtained
the effective potential.
Evaluating the effective potential in the leading order of the $1/N$
expansion, we found that the broken chiral symmetry was
restored for the sufficiently
large positive curvature in arbitrary dimensions $2 \leq D < 4$.
In the case $\lambda > \lambda_{cr}$,
the chiral symmetry is restored for a large positive curvature and the
phase transition is of the first order.
In the case $\lambda \leq \lambda_{cr}$,
the chiral symmetry is broken for any negative values of the curvature
and the second order phase transition is caused at $R=0$.
In the both cases we obtained the analytical expression of the critical
curvature $R_{cr}$ which divides the chiral symmetric and
asymmetric phases.

The large positive curvature may spoil the validity of the weak
curvature expansion.
$R_{cr}$ is, however, exactly equal to that obtained in de Sitter
space\cite{IMM} and Einstein universe\cite{IIM} at $D \rightarrow 4$
shown in Fig.4.
In four dimensions ultraviolet divergences appear in terms independent
of the curvature $R$ and linear in $R$ only and
higher order terms are ultraviolet finite.
Expanding the exact results in de Sitter space and Einstein universe
asymptotically about $R=0$,
we obtain a $R^{2}$ term of the effective potential
in a space-time with a constant positive curvature.
\begin{equation}
     V(\sigma)=V_{0}(\sigma)+V_{\sR}(\sigma)+V_{\sR2}(\sigma)
     +\mbox{O}(R^{3})\, ,
\end{equation}
where  $V_{0}(\sigma)$ and $V_{\sR}(\sigma)$ are given by
Eqs.(\ref{v0:nonren}) and (\ref{vr:nonren}) respectively,
the $R^{2}$ term $V_{\sR2}(\sigma)$ reads
\begin{equation}
     V_{\sR2}(\sigma)=-\frac{\tr \11}{(4\pi)^{\sD/2}}
     \Gamma\left(1-\frac{D}{2}\right)
     \frac{R^{2}}{5760}\frac{(D-2)(D-3)(2+5D)}{D(D-1)}
     \sigma^{\sD-4}\, .
\label{v:r2}
\end{equation}
At the four dimensional limit the $R^{2}$ term (\ref{v:r2}) reduces to
\begin{equation}
     \frac{V_{\sR2}(\sigma)}{\mu^{\sD}}
     =\frac{\tr \11}{(4\pi)^{2}}\frac{11 R^{2}}{17280}
     \left(C-\frac{173}{66}-\ln \left(\frac{\sigma}{\mu}\right)^{2}\right)\, ,
\label{r2:4d}
\end{equation}
where the divergent part $C$ is deffined in Eq.(\ref{div:c}).
The divergence in Eq.(\ref{r2:4d}) appears from the mass singularity
at $\sigma\rightarrow 0$ and the normalization condition $V(0)=0$.
Only an infrared divergence appears in the $R^{2}$ term (\ref{r2:4d}).
In the case of compact spaces the infrared divergence does not appear.
Thus the $R^{2}$ term has no contribution to the phase transition
in de Sitter space and Einstein universe.
The critical curvature is determined by the terms
involving ultraviolet divergences at four-dimensional limit
and the weak curvature expansion gives the exact result.
Therefore the weak curvature expansion is useful even for a large positive
curvature near four dimensions.

In the other dimensions it does not seem plausible
to compare the critical curvature (\ref{cr:r}) directly with
the results obtained in de Sitter space and Einstein universe, because
the global topology of the space-time may play a crucial role
for the phase transition.
To discuss the validity of the weak curvature expansion in
the other dimensions we must introduce the topological
effects in the effective potential (\ref{v:ren}).

We are interested in applying our result to critical phenomena in the early
universe.
For example we would like to study the dynamical scenario of
an inflationary universe in the four-fermion theory.
For this purpose we must calculate the stress-energy tensor
of the four-fermion theory in an expanding universe and leave it future
researches.

\subsection*{Acknowledgments}

\hspace*{\parindent}
The author would like to thank  Kennichi Ishikawa, Seiji Mukaigawa
and Taizo Muta for useful comments.
I am indebted to members of our Laboratory for
encouragements and discussions.


\begin{thebibliography}{99}
\baselineskip 18pt
\bibitem{NJL} Y.~Nambu and G.~Jona-Lasinio,
              {\it Phys. Rev.} {\bf 122}, 345 (1961).
\bibitem{TC}  S.~Weinberg, {\it Phys. Rev.} {\bf D13}, 974 (1976);
                            {\bf 19}, 1277 (1978);\\
               L.~Susskind, {\it ibid.} {\bf 20}, 2619 (1979).
\bibitem{IKM}  T.~Inagaki, T.~Kouno and T.~Muta,
               {\it Int. J. of Mod. Phys.} {\bf A10}, 2241 (1994);\\
               T.~Inagaki,
               {\it Hiroshima U. preprint}, HUPD-9527, 1995 (unpublished).
\bibitem{CURV} T.~Inagaki, T.~Muta and S.~D.~Odintsov,
               {\it Mod. Phys. Lett.} {\bf A8}, 2117 (1993).
\bibitem{CGN}  H.~Itoyama,
               {\it Prog. Theor. Phys.} {\bf 64}, 1886 (1980).
\bibitem{BK}   I.~L.~Buchbinder and E.~N.~Kirillova,
               {\it Int. J. Mod. Phys.} {\bf A4}, 143 (1989).
\bibitem{EOS}  E.~Elizalde, S.~D.~Odintsov, and Yu.~I.~Shil'nov,
               {\it Mod. Phys. Lett.} {\bf A9}, 913 (1994).
\bibitem{IMM}  T.~Inagaki, S.~Mukaigawa and T.~Muta,
               {\it Phys. Rev.} {\bf D52}, 4267 (1995).
\bibitem{ELOS} E.~Elizalde, S.~Leseduarte, S.~D.~Odintsov and Yu.~I.~Shil'nov,
               {\it U. of Barcelona preprint}, UB-ECM-PF 95/11,
               1995 (unpublished).
\bibitem{IIM}  T.~Inagaki, K.~Ishikawa and T.~Muta,
               {\it Hiroshima U. preprint}, HUPD-9517, 1995 (unpublished).
\bibitem{SACH} I.~Sachs, {\it PhD thesis}, Swiss Federal
               Institute of Technology, 1994.
\bibitem{GNRENG}T.~Muta,
                {\it Nagoya Spring School on Dynamical
                Symmetry Breaking,} ed. K.\\Yamawaki (World
                Scientific, 1992);
                H.-J.~He, Y.-P~Kuang, Q.~Wang and Y.-P.~Yi,
		{\it Phys. Rev.} {\bf D45}, 4610 (1992).
\bibitem{MTW}  C.~W.~Misner, K.~S.~Thorne and J.~A.~Wheeler,
               {\it Gravitation} (W.~H.~Freeman and Co., 1973).
\bibitem{GN}  D.~J.~Gross and A.~Neveu,
              {\it Phys. Rev.} {\bf D10}, 3235 (1974).
\bibitem{SP}   J.~Schwinger, {\it Phys. Rev.} {\bf 82}, 664 (1951).
\bibitem{PT}    L.~Parker and D.~J.~Toms,
                   {\it Phys. Rev.} {\bf D29}, 1584 (1984).
\bibitem{RNC}  A.~Z.~Petrov, {\it Einstein Spaces}
               (Pergamon, Oxford, 1969).
\end{thebibliography}
\end{document}